%%%%%%%%%%%%%%%%%%%%%%%%%%%%%%%%%%%%%%%%%%%%%%%%%%%%%%%%%%%%%%%%%%%%%%
\input harvmac

\input epsf
\noblackbox

%%%%% FONTS
\def\inbar{\,\vrule height1.5ex width.4pt depth0pt}
\font\cmss=cmss10 \font\cmsss=cmss10 at 7pt
\def\IZ{\relax\ifmmode\mathchoice
{\hbox{\cmss Z\kern-.4em Z}}{\hbox{\cmss Z\kern-.4em Z}}
{\lower.9pt\hbox{\cmsss Z\kern-.4em Z}}
{\lower1.2pt\hbox{\cmsss Z\kern-.4em Z}}\else{\cmss Z\kern-.4em
Z}\fi}
\def\IB{\relax{\rm I\kern-.18em B}}
\def\IC{{\relax\hbox{$\inbar\kern-.3em{\rm C}$}}}
\def\ID{\relax{\rm I\kern-.18em D}}
\def\IE{\relax{\rm I\kern-.18em E}}
\def\IF{\relax{\rm I\kern-.18em F}}
\def\IG{\relax\hbox{$\inbar\kern-.3em{\rm G}$}}
\def\IGa{\relax\hbox{${\rm I}\kern-.18em\Gamma$}}
\def\IH{\relax{\rm I\kern-.18em H}}
\def\II{\relax{\rm I\kern-.18em I}}
\def\IK{\relax{\rm I\kern-.18em K}}
\def\IP{\relax{\rm I\kern-.18em P}}

\font\cmss=cmss10 \font\cmsss=cmss10 at 7pt
\def\IR{\relax{\rm I\kern-.18em R}}

\def\sslash#1{#1\hskip-0.5em /}

%%%%%REFERENCES

\lref\classcon{P. Green and T. Hubsch,
``Phase Transitions Among (Many Of) Calabi-Yau Compactifications,''
Phys.\ Rev.\ Lett.\  {\bf 61}, 1163 (1988);
P.~Candelas, P.~S.~Green and T.~Hubsch,
``Finite Distances Between Distinct Calabi-Yau Vacua: 
(Other Worlds Are Just Around The Corner),''
Phys.\ Rev.\ Lett.\  {\bf 62}, 1956 (1989). }
\lref\ds{M. Dine and E. Silverstein, ``New M-theory Vacua
with Frozen Moduli'', hep-th/9712166;
J. Harvey, S. Kachru, G. Moore, and E. Silverstein,
``Asymmetric D-factory'', may appear;
I. Brunner, A. Rajaraman, and M. Rozali, 
``D-branes on Asymmetric Orbifolds'', hep-th/9905024.}
\lref\GP{E. Gimon and J. Polchinski,
``Consistency Conditions for Orientifolds and D-Manifolds,''
Phys.\ Rev.\  {\bf D54}, 1667 (1996),
hep-th/9601038. }
\lref\dougmoore{M. Douglas and G. Moore, 
``D-branes, Quivers, and ALE Instantons,''
hep-th/9603167.}
\lref\kks{S. Kachru, J. Kumar, and E. Silverstein, 
`Vacuum energy cancellation in a non-supersymmetric string,''
Phys.\ Rev.\  {\bf D59}, 106004 (1999), hep-th/9807076.}
\lref\zam{A.B. Zamolodchikov, ``Conformal Symmetry and
Multicritical Points in Two-Dimensional Quantum Field
Theory'', Sov. J. Nucl. Phys. {\bf 44} 530 (1986).}
\lref\BFSS{T. Banks, W. Fischler, S. Shenker, and L. Susskind,
``M theory as a matrix model: A conjecture,''
Phys.\ Rev.\  {\bf D55}, 5112 (1997),
hep-th/9610043.}
\lref\cpn{E.~Witten,
``Instantons, The Quark Model, And The 1/N Expansion,''
Nucl.\ Phys.\  {\bf B149}, 285 (1979);
A.~D'Adda, A.~C.~Davis, P.~Di Vecchia and P.~Salomonson,
``An Effective Action For The Supersymmetric Cp**(N-1) Model,''
Nucl.\ Phys.\  {\bf B222}, 45 (1983).
}
\lref\LG{B.~R.~Greene, C.~Vafa and N.~P.~Warner,
``Calabi-Yau Manifolds And Renormalization Group Flows,''
Nucl.\ Phys.\  {\bf B324}, 371 (1989); 
E. Martinec, "Criticality, Catastrophes and Compactifications"
in
Brink, L. et al editors, Physics and mathematics of
strings, 389-433 (V. Knizhnik memorial volume)
}
\lref\phases{E. Witten, ``Phases of N=2 Theories in Two Dimensions''
Nucl.\ Phys.\  {\bf B403}, 159 (1993),
hep-th/9301042.}
\lref\ksorb{S. Kachru and E. Silverstein, ``4d Conformal
Theories and Strings on Orbifolds''
Phys.\ Rev.\ Lett.\  {\bf 80}, 4855 (1998), hep-th/9802183.}
\lref\lnv{A. Lawrence, N. Nekrasov, and C. Vafa,
``On conformal field theories in four dimensions,''
Nucl.\ Phys.\  {\bf B533}, 199 (1998),
hep-th/9803015.}
\lref\bkv{M. Bershadsky, Z. Kakushadze, and C. Vafa,
``String expansion as large N expansion of gauge theories,''
Nucl.\ Phys.\  {\bf B523}, 59 (1998),
hep-th/9803076; M. Bershadsky
and A. Johansen, ``Large N limit of orbifold field theories,''
Nucl.\ Phys.\  {\bf B536}, 141 (1998),
hep-th/9803249}
\lref\AdSCFT{J. Maldacena, 
``The large-N limit of superconformal field theories and supergravity,''
Adv.\ Theor.\ Math.\ Phys.\  {\bf 2}, 231 (1998)
hep-th/9711200; S. Gubser, I. Klebanov, and A. Polyakov,
``Gauge theory correlators from non-critical string theory,''
Phys.\ Lett.\  {\bf B428}, 105 (1998),
hep-th/9802109;
E. Witten, ``Anti-de Sitter space and holography,''
Adv.\ Theor.\ Math.\ Phys.\  {\bf 2}, 253 (1998),
hep-th/9802150.}
\lref\frampton{P. H. Frampton,
 ``AdS/CFT string duality and conformal gauge field theories,''
Phys.\ Rev.\  {\bf D60}, 041901 (1999),
hep-th/9812117.
}
\lref\terning{C.~Csaki, W.~Skiba, and J. Terning, 
``Beta functions of orbifold theories and the hierarchy problem,''
hep-th/9906057.}
\lref\adk{I. Antoniadis, J. Derendinger and C. Kounnas, ``Nonperturbative
Temperature Instabilities in N=4 Strings,'' Nucl. Phys. {\bf B551} (1999)
41, hep-th/9902032.}
\lref\thooft{G. 't Hooft, in {\it Recent Developments in Gauge Theories}
G.~'t~Hooft, C.~Itzykson, A.~Jaffe, H.~Lehmann, P.~K.~Mitter, 
 I.~M.~Singer, and R.~Stora, eds. (Plenum Press, New York, 1980).}
\lref\vafaquantum{C. Vafa,
``Quantum Symmetries Of String Vacua,''
Mod.\ Phys.\ Lett.\  {\bf A4}, 1615 (1989).}
\lref\IMSY{N.~Itzhaki, J.~M.~Maldacena, J.~Sonnenschein and S.~Yankielowicz,
``Supergravity and the large N limit of theories with sixteen  supercharges,''
Phys.\ Rev.\  {\bf D58}, 046004 (1998),
hep-th/9802042.}
\lref\matrixstring{L. Motl, 
``Proposals on nonperturbative superstring interactions,''
hep-th/9701025;
R.~Dijkgraaf, E.~Verlinde and H.~Verlinde,
``Matrix string theory,''
Nucl.\ Phys.\  {\bf B500}, 43 (1997),
hep-th/9703030; T.~Banks and N.~Seiberg,
``Strings from matrices,''
Nucl.\ Phys.\  {\bf B497}, 41 (1997),
hep-th/9702187.}
\lref\silvwitt{E. Silverstein and E. Witten, ``Criteria for
Conformal Invariance of (0,2) Models'',
Nucl.\ Phys.\  {\bf B444}, 161 (1995),
hep-th/9503212.}
\lref\conifold{A. Strominger, 
``Massless black holes and conifolds in string theory,''
Nucl.\ Phys.\  {\bf B451}, 96 (1995),
hep-th/9504090; 
S. Kachru, N. Seiberg, and E. Silverstein,
``SUSY Gauge Dynamics and Singularities of 4d N=1 String Vacua,''
Nucl.\ Phys.\  {\bf B480}, 170 (1996),
hep-th/9605036.}
\lref\ade{E. Witten, ``String Theory
Dynamics in Various Dimensions'', 
Nucl.\ Phys.\  {\bf B443}, 85 (1995),
hep-th/9503124.}
\lref\donedfivecft{E. Witten, ``On the Conformal Field Theory of
the Higgs Branch'', 
JHEP {\bf 9707}, 003 (1997),
hep-th/9707093; 
N.~Seiberg and E.~Witten,
``The D1/D5 system and singular CFT,''
JHEP {\bf 9904}, 017 (1999), 
hep-th/9903224; O. Aharony and M. Berkooz, 
``IR dynamics of D = 2, N = (4,4) 
gauge theories and DLCQ of 'little  string theories',''
JHEP {\bf 9910}, 030 (1999)
hep-th/9909101.}
\lref\wittenD{E. Witten, ``Bound States of Strings and p-branes'',
Nucl.\ Phys.\  {\bf B460}, 335 (1996),
hep-th/9510135.}
\lref\vafa{C. Vafa, private discussion.}
\lref\iz{
R.~Iengo and C.~Zhu,
``Evidence for nonvanishing cosmological 
constant in nonSUSY superstring  models,''
hep-th/9912074.}
%\lref\ab{O. Aharony and M. Berkooz,}
\lref\doug{M. Douglas, 
``Enhanced gauge symmetry in M(atrix) theory,''
JHEP {\bf 9707}, 004 (1997),
hep-th/9612126.}
%\lref\paulstud{P. Aspinwall and S.?, in progress.}
\lref\paul{P. Aspinwall,
``Resolution of Orbifold Singularities in String Theory'',
To appear in 'Essays on Mirror Manifolds 2'. 
In Greene, B. (ed.), Yau, S.T. (ed.): {\it Mirror symmetry II} 355-379,
hep-th/9403123.}  
\lref\kkstach{S. Kachru, J. Kumar, and E. Silverstein,
``Orientifolds, RG flows, and closed string tachyons,''
hep-th/9907038.}
\lref\oldjoe{S.P. de Alwis, J. Polchinski, and R. Schimmrigk, 
``Heterotic Strings with Tree Level Cosmological Constant,''
Phys. Lett. {\bf B218} (1989) 449.}
\lref\oldpot{See for instance:
 V.A.~Kostelecky and S.~Samuel,
``The Tachyon Potential In String Theory,''
{\it Presented at 1988 Mtg. of Div. of Particle and Fields of the APS,
                  Storrs, CT, Aug 15-18, 1988}\semi 
T. Banks, ``The Tachyon Potential in String Theory,'' Nucl. Phys.
{\bf B361} (1991) 166\semi   
A. Belopolsky and B. Zwiebach, 
``Off Shell Closed String Amplitudes: Towards a Computation of
the Tachyon Potential,'' Nucl. Phys. {\bf B442} (1995) 494, hep-th/9409015.} 
\Title{\vbox{\baselineskip12pt\hbox{hep-th/9912244}
\hbox{IASSNS-HEP-99/119, SLAC-PUB-, SU-ITP-99/54}
}}
{\vbox{\centerline{On the Critical Behavior of D1-brane Theories}
}
}                                   

\centerline{
Eva Silverstein$^{a,b}$ and Yun S. Song$^{b}$ 
\foot{{\tt evas@slac.stanford.edu, yss@leland.stanford.edu}}
 }
\bigskip 
\centerline{$^{a}$ School of Natural Sciences}
\centerline{~~~Institute for Advanced Study}
\centerline{~~~Olden Lane}
\centerline{~~~Princeton, NJ 08540}
\medskip
\centerline{$^{b}$ Department of Physics and SLAC}
\centerline{~~~Stanford University}
\centerline{~~~Stanford, CA 94305/94309}
\bigskip
\noindent

We study renormalization-group flow patterns in theories arising
on  D1-branes in various supersymmetry-breaking backgrounds.
We argue that the theory of N D1-branes 
transverse to an orbifold space 
can be fine-tuned to flow to the corresponding
orbifold conformal field theory in the infrared,
for particular values of the couplings and
theta angles which we determine using the
discrete symmetries of the model.  
By calculating various nonplanar contributions
to the scalar potential in the worldvolume theory, we show
that fine-tuning is in fact required at finite N, as would
be generically expected.  We further comment on the presence
of singular conformal field theories (such as those whose
target space includes a ``throat'' described by an exactly
solvable CFT) in the non-supersymmetric context.  Throughout
the analysis two applications are considered:  to 
gauge theory/gravity duality and to linear sigma model
techniques for studying worldsheet string theory.

\Date{December 1999}
%\draftmode

\newsec{Introduction}

Getting a handle on the quantum behavior of non-supersymmetric
systems is important in order to be able to concretely approach 
hierarchy problems 
and other aspects of supersymmetry breaking.
This is also ultimately important in studying any theory, 
supersymmetric or not,
since generic phenomena are
not protected by supersymmetry.

In this paper we will analyze the quantum field theories
arising on D1-branes in various supersymmetry-breaking
backgrounds.  We will focus mostly on D1-branes transverse
to symmetric orbifold singularities, but will for
comparison consider also D1-branes in combination with
D5-branes and transverse to asymmetric orbifold backgrounds.

We have in mind two main motivations (or potential applications
for these theories).  The first concerns their 
role in gravity/gauge theory dualities
\BFSS\AdSCFT.  The strong coupling
regime in the RG flow of the worldvolume
theory of D1-branes in flat space has
a weakly-coupled supergravity dual \IMSY.  On a circle, 
this theory defines matrix string theory \matrixstring;
there the strong coupling regime of the RG flow
provides a non-perturbative
formulation of interacting string theory.    
It would be very nice to know whether this flow
pattern persists in situations with less supersymmetry.
In particular we would like to understand
whether the regime of finite (but large)
Yang-Mills gauge coupling survives, whether a mass gap
develops, and so on.  As in \ksorb, one can use
the gravity/gauge theory duality to translate
the question of whether a string-scale vacuum
energy arises into a potentially simpler question
in the dual gauge theory.  The second application is to
linear sigma model techniques \zam\LG\phases\ for studying
worldsheet conformal field theory via a
relatively simple ultraviolet
field theory that flows to the desired 
CFT in the infrared.  

As we will review, in the large-N
limit this flow pattern does persist for ``quiver''
gauge theories \dougmoore\GP\ obtained on D1-branes
at orbifold singularities \ksorb\lnv\bkv.  Even for finite N, these
theories have the promising feature that the 
target space (classical moduli space) of the theory in
the ultraviolet would, if taken as the target space
of a sigma model, produce a conformal field theory
(here simply the corresponding
orbifold conformal field theory).  This is quite
different from the situation obtained in studying, for example,
the  $\IC\IP^n$ model, for which the classical
moduli space is not Ricci-flat, and in which
a large-N analysis reveals the presence of a mass
gap \cpn.  

Because of the promising start the quiver theories
have at leading order 
in the 1/N expansion, it is worth checking explicitly
whether these pleasant features persist at finite N.
We find by explicit calculation that they do not
for generic $(1+1)d$ quiver theories.  (Similar questions
in $(3+1)d$ were studied by explicit calculation in
\frampton\terning).  
We show, by way of constrast, that they do persist,
at the order we compute,
for the worldvolume theories arising on D1-branes
transverse to asymmetric orbifolds such as \kks.

We argue, using various simple features of this 
type of model (such
as the form of the UV target space mentioned above),
that the quiver gauge theories can however be fine-tuned to
flow to the corresponding orbifold conformal field
theory for particular choices of the couplings and
$\theta$-angles.  In particular we 
show how to identify using the
quiver theories the values of the theta angles at
the orbifold conformal field theory point 
for $\IZ_k$ orbifolds with $k>2$ (as determined
using wrapped D-branes in \doug, giving a generalization
of the result for $\IZ_2$ orbifolds determined
by Aspinwall \paul).

This statement about the flow involves a comparison of relevant
operators and discrete symmetries in the two limiting
theories.  The situation might be compared to that of
non-supersymmetric Landau-Ginzburg theories which flow to minimal
model CFT's in the IR \zam.  It has potential
utility as an approach for studying twisted-sector
tachyon condensation in perturbative string theory.
The fact that there is a field theory flow between
the two limiting theories raises the interesting
question as to whether there is a gravitational
dual to this trajectory in coupling space.

In supersymmetric contexts, one often finds subspaces
on the moduli space of $2d$ conformal field theories
on which the CFT becomes singular.  (Examples are
the points of non-perturbatively enhanced gauge
symmetry in type IIA compactification on K3 \ade,
the conifold singularities of Calabi-Yau compactification
\classcon\phases, and the small instanton singularities
of the D1-D5 system \donedfivecft.)
We study a simple model (based trivially on a CFT that factorizes
into a supersymmetric part times a nonsupersymmetric part)
in which this persists without supersymmetry and discuss
the potential implications for the spacetime string background
described by this worldsheet conformal field theory.

The outline of this paper is as follows.  In \S2\ we introduce
the theories we will study.  In \S3\ we discuss the symmetry
structure, operator content, and target space geometry of
the quiver Yang-Mills theories (with particularly symmetric
choices for the parameters of the gauge theories) 
and compare them to those of
the corresponding orbifold CFT.  The agreement
between the two provides strong evidence
that the quiver theories flow (upon appropriate fine-tuning) to
the corresponding orbifold CFT in the infrared, as we
explain in \S3.3; we discuss interesting applications of
this fact in \S3.4.  In \S4\ we present calculations
which show
that  fine-tuning is in fact required at subleading orders in
the $1/N$ expansion of the non-supersymmetric
quiver theories; we constrast this to the case
of D-branes at asymmetric orbifolds where such
fine-tuning is not required at the order we
compute.  Finally, in \S5\ we discuss a simple example
of a non-supersymmetric string background in which the
worldsheet CFT becomes singular and consider the
worldvolume theories on D-branes in such backgrounds.   
 
\newsec{The Theories}

\smallskip

In this section we present the theories we will study.
Most of our analysis will involve the symmetric orbifold
theories in \S2.1, which we will compare and contrast
with analogous results for the theories in \S2.2\ and \S2.3.

\subsec{Quiver Theory}

The worldvolume theories on Dp-branes at symmetric orbifold
singularities were worked out in \GP\dougmoore.  
In general, one can consider an orbifold $\IC^4/\Gamma$
where $\Gamma$ is a subgroup of the Poincare group 
of $\IC^4$.  The orbifold
group acts on the Chan-Paton factors on the open string
endpoints in addition to acting on the Lorentz quantum
numbers of the fields.  In the case that $\Gamma=\IZ_k$,
one finds a gauge group $U(N)^k$ and matter in
various bifundamental representations determined by
the details of the geometrical action of $\Gamma$.
If the Chan-Paton factors sit in
a regular representation of $\Gamma$, then the quiver
theory describes Dp-branes that can all move off the orbifold
fixed point.  In the case of a non-regular representation,
there are fractional branes stuck at the fixed point, which
at least in some cases have been seen to be equivalent
to higher dimensional D-branes wrapped around a collapsed
cycle implicit in the singularity.       

We will consider the following supersymmetry-breaking
model in some detail.  We will take a D1-brane transverse
to $\IC/\IZ_3\times \IC^3$.  The $\IZ_3$ acts on the 
four complex coordinates $(x^1,x^2,x^3,x^4)$ transverse
to the D1-brane by 
\eqn\action{(x^1,x^2,x^3,x^4)\to 
(e^{2\pi i({2\over 3})}x^1, x^2, x^3, x^4).
}
The unorbifolded D1-brane has a gauge group
$U(3N)$ with fermions and scalars in the adjoint
representation.  Left-moving fermions 
$\chi_L^\alpha, \alpha=1,\dots,8$ transform in the
${\bf 8_s}$ of the $SO(8)$ rotation group transverse
to the branes, and right-moving fermions 
$\chi_R^{\dot\alpha}, \dot\alpha=1,\dots,8$
transform in the ${\bf 8_c}$
of that $SO(8)$.  Breaking up the eight real left-moving fermions
into four complex fermions, one finds that the action
\action\ rotates them by $e^{{{2\pi i}\over 3}}$, and
similarly for the right-movers.  This means in particular
that the orbifold breaks all the supersymmetry.  
The action on the scalars
$(X^1, X^2, X^3, X^4)$, whose eigenvalues parameterize
the position of the brane on the transverse $\IC^4$, is
as indicated in \action.    

Projecting onto invariant states (taking the regular 
representation for the action on the Chan-Paton indices),
on finds the following spectrum. The gauge group is $U(N)^3$.  The 
components of $X^2, X^3$, and $X^4$ that are invariant
under the orbifold transform in the adjoint of $U(N)^3$.
$X^1$ and the fermions $\chi_L, \chi_R$ transform in
the $(N,\overline{N}, 1)+(1,N,\overline{N})+(\overline{N}, 1, N)$
representation.  
The quiver diagram for this theory \dougmoore\ is 
shown in Figure 1.
%%{\it !!Z3 Model Quiver Diagram; include conjugates?}

The interactions of the theory are those of the 
unorbifolded ${\cal N}=8$ theory which involve 
worldvolume fields invariant under the orbifold.  
The diagonal $U(1)$ in the $U(N)^3$ gauge group is
totally decoupled (nothing is charged under it).  There
are two ``relative'' $U(1)$ factors which couple.  The
Lagrangian includes theta-terms $\theta_i\int F_i$
where $F_i$ are the field
strengths of the three $U(1)$ factors.  We will
analyze the symmetry structure of the theory arising from
these interactions (choosing particularly symmetric
values for the quiver theory couplings) in studying
their renormalization-group flow in \S3.

%%%%%%%%%%%%%%%%%%%%%%% Figure 1 %%%%%%%%%%%%%%%%%%%%%%%%%%%%%%%%%
\bigskip
\centerline{\vbox{\hsize=4.5in\tenpoint 
\centerline{\epsfbox{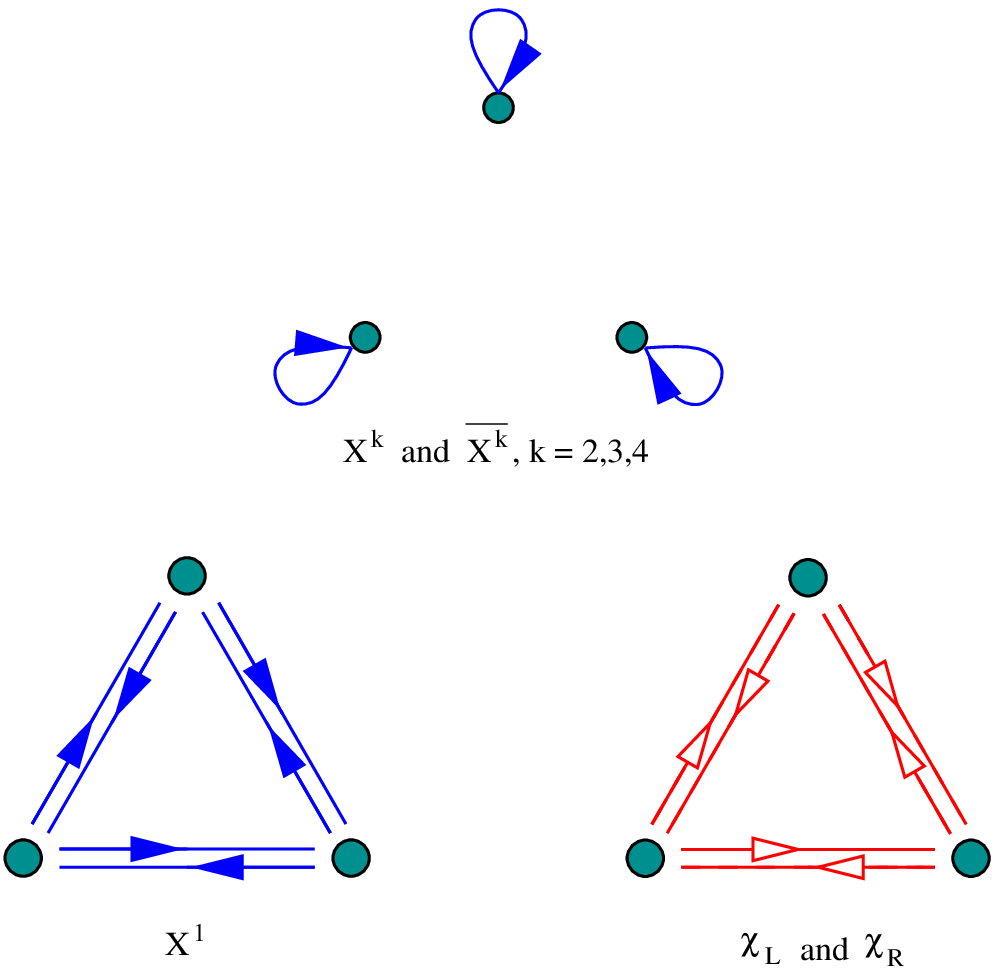}}
\smallskip\noindent 
\centerline{ Figure 1: $\IC/\IZ_3$ model quiver diagram.}}}
\bigskip
%%%%%%%%%%%%%%%%%%%%%%%%%%%%%%%%%%%%%%%%%%%%%%%%%%%%%%%%%%%%%%%%%%

\subsec{D1--D5}

We will also briefly consider supersymmetry-breaking theories
involving both D1-branes and D5-branes.  Let us consider
a $\IC/\IZ_3$ orbifold transverse to the D1--D5 system, taking the
regular representation for the action of the orbifold
group on both the D1-branes and the D5-branes.  (Otherwise
the (1--5) and (5--1)-sector strings would be projected
out by the orbifold action.)  This involves considering
3 sets of $Q_1$ D1-branes and 3 sets of $Q_5$ D5-branes.
Before orbifolding, the
(1--1) sector strings are as described in the previous
subsection, and the (1--5) and (5--1) strings form hypermultiplets
in the ${\bf (3Q_1,3Q_5)}$ representation of the 
$U(3Q_1)\times U(3Q_5)$ symmetry group of the D1-branes and
D5-branes.  After orbifolding, we obtain:
%%{\it !!!D1-D5 quiver diagram}     

%%%%%%%%%%%%%%%%%%%%%%%% Figure 2%%%%%%%%%%%%%%%%%%%%%%%%%%%%%%%
\bigskip
\centerline{\vbox{\hsize=4.5in\tenpoint 
\centerline{\epsfbox{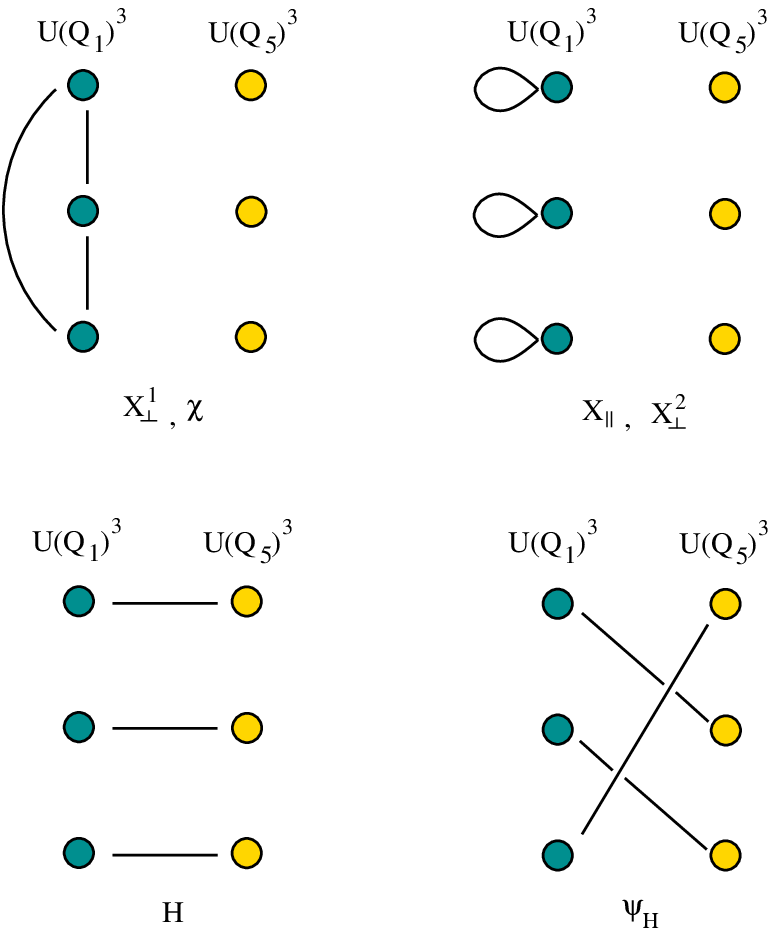}} 
\noindent
Figure 2: D1--D5 quiver diagrams.
We schematically denote conjugate pairs by a single line, and
therefore omit arrows in the diagram.  $X_\perp$ are the coordinates
perpendicular to the D1--D5 brane system, whereas $X_\parallel$ denote
the position of D1-branes on D5-branes. $H$ and $\psi_H$ are
hypermultiplet fields coming from the (1--5) and (5--1) strings.}}
\bigskip
%%%%%%%%%%%%%%%%%%%%%%%%%%%%%%%%%%%%%%%%%%%%%%%%%%%%%%%%%%%%%%%%%%

\subsec{Asymmetric Orbifolds}

We will find in some sense more positive results for the
case of D-branes on certain asymmetric orbifold backgrounds \kks.
In this subsection we briefly review what we will need about
their worldvolume theories \ds.  Consider an
asymmetric orbifold involving an action by $(-1,1)^4(-1)^{F_R}$
(that is, a $\IZ_2$ reflection of the left-moving coordinates
and no action on the right-moving coordinates on the string
worldsheet).  This is a symmetry  
of $T^4$ compactification at the $SO(8)$-symmetric point in Narain
moduli space, and is a basic ingredient in the models considered
in \kks.  

This action maps D1-branes transverse to the $T^4$ into
D5-branes wrapped on it, and therefore is only a symmetry
if we take $Q_1=Q_5$ in the unorbifolded theory.  The
orbifold action then acts off-diagonally on (a natural
basis of) the open-string Hilbert space.  Strings from
the (1--1) sector map to (5--5) strings, (1--5) to (5--1).    
This means that if we start with a (1--1) sector string
state $\left| \phi \right>_{1-1}$, it will automatically combine with a (5--5)
string state $\left| g\phi \right>_{5-5}$ (where $g\phi$  denotes the
result of acting with the orbifold group on the worldsheet modes $\phi$)
to form an invariant linear combination 
$\left|\phi\right>_{1-1}+\left|g\phi\right>_{5-5}$.  This will be significant
later.

\newsec{General Features}

In this section we will assemble information on the
target space, symmetries, and operator content of
the quiver gauge theories.  This will provide strong
evidence for a (possibly fine-tuned) flow from the quiver
Yang-Mills theories for particular values of the
parameters to orbifold conformal
field theory in the infrared.  In particular in
\S3.1\ we will determine the appropriate values of
the $\theta$-parameters corresponding to the
orbifold conformal field theory.    

\subsec{Interactions and Symmetries}

We are interested in whether the quiver gauge theory
obtained on D1-branes on an orbifold background flows
to the corresponding orbifold CFT in the infrared.  
A $\IZ_k$ orbifold CFT has two types of discrete symmetries.
The first is an exchange of the $g$-twisted sector with
the $g^{-1}$-twisted sector (for all orbifold group
elements $g$) combined with left$\leftrightarrow$right 
exchange (parity).  Let us call this transformation 
``t-Parity''.    
The second is
the $\IZ_k$ quantum symmetry \vafaquantum\ which
constrains the correlators of twisted sector
vertex operators.  
Also,
the orbifold conformal field theory is nonsingular.
In supersymmetric contexts, this requires that
the UV Yang-Mills theory not have a Coulomb branch.
  
For simplicity we will here consider $\IZ_3$ orbifolds such as
the $\IC/\IZ_3$ case that is our main example, but the
results of this subsection are more general (and in
particular apply to the supersymmetric cases of
\paul\doug).
If we choose
the gauge couplings and matter self-interactions as inherited
from the unorbifolded theory, then these terms respect
t-parity and a $\IZ_3$ symmetry under which the
three $U(N)$ factors in the gauge group are cyclically
permuted.  

Let us consider the $\theta$-parameters in the field theory.
In general these couple via the terms
\eqn\gentheta{
{\cal L}_\theta=\theta_1\int F_1+\theta_2\int F_2
+\theta_3\int F_3
}
This can be rewritten as (writing $\alpha=e^{{{2\pi i}\over 3}}$)
\eqn\complbas{\eqalign{{\cal L}_\theta
&= \int ({{\theta_1+\alpha^{-1}\theta_2+\alpha\theta_3}\over 3})
(F_1+\alpha F_2+\alpha^{-1}F_3)
+({{\theta_1+\alpha\theta_2+\alpha^{-1}\theta_3}\over 3})
(F_1+\alpha^{-1} F_2+\alpha F_3) \cr
& ({{\theta_1+\theta_2+\theta_3}\over 3})
(F_1+F_2+F_3)\cr
}
}
This basis is useful for studying the possible correspondence with
orbifold conformal field theory because each term here
is an eigenstate of the $\IZ_3$ symmetry which permutes the
three $U(1)$ factors.  
Let us set
\eqn\redef{\eqalign{
&\eta={{\theta_1+\alpha^{-1}\theta_2+\alpha\theta_3}\over 3}\cr
&\bar\eta={{\theta_1+\alpha\theta_2+\alpha^{-1}\theta_3}\over 3}\cr
}
}
The theory is invariant under the shifts $\theta_i\to\theta+2\pi n_i$.
for integer $n_i$.  In terms of the complex parameter $\eta$, this is
the equivalence
\eqn\shiftsymm{
\eta\cong\eta+{{2\pi}\over 3}(n_1+\alpha^{-1}n_2+\alpha n_3).
}

We would like to find the values $\eta$ can take such that:

\noindent (i)  There is a $\IZ_3$ symmetry permuting the
three $U(1)$ factors,

\noindent (ii) There is a t-Parity symmetry, and

\noindent (iii) The model has no Coulomb branch.

Translating these conditions into conditions on $\eta$, they
become 
\eqn\conditions{\eqalign{
&(i)~~\eta+{{2\pi}\over 3}(m_1+\alpha^{-1}m_2+\alpha m_3)=\alpha\eta\cr
&(ii)~~\eta+{{2\pi}\over 3}(n_1+\alpha^{-1}n_2+\alpha n_3)=-\bar\eta\cr
&(iii)~~\eta\not\cong 0 \cr
}
}
The condition (iii) must be imposed since
it ensures that there is a nonzero background electric field
which contributes positive vacuum energy obstructing
the Coulomb branch,
as explained for this type of model in \phases.

These conditions are solved by 
\eqn\ans{
\eta={1\over 3}\biggl[-\pi+i{\sqrt{3}\over 2}
\biggl({{2\pi}\over 3}\biggr)\biggr].
}

%This can be rewritten as
%\eqn\rewrite{\eqalign{
%&{\cal L}_\theta=\int
%({{\theta_1-\theta_2}\over 3})(F_1-F_2) +
%({{\theta_2-\theta_3}\over 3})(F_2-F_3) +
%({{\theta_3-\theta_1}\over 3})(F_3-F_1))\cr
%&+ ({{\theta_1+\theta_2+\theta_3}\over 3})
%(F_1+F_2+F_3)\cr
%}}

Another simple basis to work in is the following.
\eqn\simple{
{\cal L}_\theta=\int
({{2\theta_1-\theta_2-\theta_3}\over 3})(F_1-F_2)+
({{2\theta_3-\theta_1-\theta_2}\over 3})(F_3-F_2)
+\theta_{tot}F_{tot}
}
where $\theta_{tot}F_{tot}=({{\theta_1+\theta_2+\theta_3}\over 3})
(F_1+F_2+F_3)$.  
%The physics is unchanged under $2\pi$ shifts
%of the $\theta_i$.   
Let us set $\kappa_{12}\equiv {{2\theta_1-\theta_2-\theta_3}\over 3}$
and $\kappa_{32}\equiv {{2\theta_3-\theta_1-\theta_2}\over 3}$.
%there is correspondingly an equivalence
%\eqn\shifts{
%(\kappa_{12},\kappa_{32})
%\cong 
%(\kappa_{12}+{{2\pi}\over 3}(2n_1-n_2-n_3),
%\kappa_{32}+{{2\pi}\over 3}(2n_3-n_1-n_2))
%} 

%In order to have a parity symmetry in the $\theta$ sector,
%we need 
%\eqn\parityreq{
%(\kappa_{12},\kappa_{32})\cong (-\kappa_{12},-\kappa_{32})
%}
%Under the $\IZ_3$ that cyclically
%permutes the $U(N)$,
%\eqn\zthreetransf{\eqalign{
%&{\cal L}_\theta \to \int
%\kappa_{12}(F_2-F_3)+\kappa_{32}(F_3-F_1)\cr
%&= -\kappa_{32}(F_1-F_2)+
%(\kappa_{32}-\kappa_{12})(F_3-F_2)\cr}
%}
%So in order to have a symmetry under this $\IZ_3$, we need
%\eqn\zthreereq{
%(\kappa_{12},\kappa_{32})\cong
%(-\kappa_{32}, \kappa_{32}-\kappa_{12})
%}

%Both \parityreq\ and \zthreereq\ are satisfied by choosing
%\eqn\choices{
%}

In this basis, the solution \ans\ for $\eta$ is
\eqn\answerkappa{
(\kappa_{12},\kappa_{32})
=(-2\pi/3, 2\pi/3)
}
This agrees with the results of \doug.  This procedure of
enforcing t-Parity, the quantum symmetry, and the absence
of a Coulomb branch of course applies much more generally
than just to the $\IZ_3$ orbifold of interest here.

%%%%%%%%%%%%%%%%%%%
%  Text for chiral symmetries
%%%%%%%%%%%%%%%%%%%

We can also consider continuous global symmetries in
addition to the discrete symmetries we have so far considered.
Chiral symmetries in particular can provide
exact, non-perturbative constraints
on renormalization group flow.
As proposed by 't Hooft \thooft, anomaly coefficients remain
invariant under the renormalization-group flow, and hence their
matching can serve as a useful consistency check.  In our  $\IC/\IZ_3$
example, however, we find by studying the
surviving interactions in the theory that
there are no chiral symmetries in the UV
with nontrivial 't Hooft anomalies.  This was noted in
the case of D1-branes in flat space in \wittenD.  In
quiver theories with for example (2,2) supersymmetry, one
does find chiral symmetries, but they act on scalars
which parameterize the target space of the D1-brane
worldvolume theory.  These symmetries therefore do not match on
to the R-symmetries in the affine Lie algebra of the
infrared superconformal field theory.  In our 
supersymmetry-breaking example, we do not find even
this type of chiral symmetry.  We
therefore cannot use the 't Hooft anomaly matching condition to
constrain the flow.  The absence of such chiral symmetries
has implications for matrix string theory as we discuss
further below.

\subsec{Moduli Space}

Let us consider the $\IC/\IZ_3$ theory of \S2.1.  The classical
moduli space of the ultraviolet Yang-Mills theory on the
D1-branes has two branches.  One of them, the Higgs branch, is
simply the orbifold space
\eqn\higgsmod{
{\cal M}^{{\rm class}}_{Higgs}=(\IC/\IZ_3\times \IR^6)^N/S_N
}
This is true simply by construction; this branch describes the
$N$ branes moving around on the orbifold space we started with.

If the branes coalesce at the singularity, they can
separate in the $\IR^6$ transverse to the orbifolded dimensions.
This yields a classical Coulomb branch of the form 
\eqn\coulmod{
{\cal M}^{\rm class}_{Coulomb}=(\IR^6)^{3N}/S_{3N}
}
This branch only exists for $\kappa_{12}=0,\kappa_{32}=0$.

\subsec{Operator Spectrum and the Flow Conjecture}

Let us summarize the situation so far.
If we consider the theory at the values of the 
$\theta$ parameters determined in \S3.1, then the
theory has several features which will be of interest:

\noindent (1) an unbroken t-Parity symmetry, and no Coulomb
branch.
  
\noindent (2) The Higgs branch \higgsmod\ is a simple
orbifold space.  

\noindent (3) There is also a $\IZ_3$ discrete symmetry
in the model, acting on operators which are gauge-invariant
in the orbifold but would not have been in the original
theory.  
      
These features (1)-(3) are also true of the orbifold
conformal field theory on $(\IC/\IZ_3\times \IR^6)^N/S_N$:

\noindent (1)  This is a 
(nonsingular) symmetric orbifold conformal
field theory, so there is a left-right
exchange symmetry combined with exchange of congugate
twisted sectors (t-Parity).  

\noindent (2) The target space is this orbifold
space.  

\noindent (3) The orbifold CFT includes twisted sectors that
transform under a discrete 
$\IZ_3$ ``quantum'' symmetry \vafaquantum\
that constrains their correlators.    

These features suggest that the ultraviolet Yang-Mills
theory might flow in the infrared to the corresponding
orbifold conformal field theory.  Let us consider the
operator spectrum of the two theories.

The orbifold CFT has relevant operators of the form
\eqn\untwisted{\eqalign{
&e^{ikY}+(images~~under~~\IZ_3~~and~~S_N~~actions)\cr 
&{1\over 2}k^2<1\cr
}
}
where $Y$ is a coordinate in $(\IC\times \IR^6)^N$.  If we consider
a compact target space then there is a finite
number of such operators.  These operators are not
charged under any symmetry of the theory.  
If we add them to the Lagrangian
they appear as contributions to the scalar potential.
Being relevant, adding them to the Lagrangian drives
the theory away from the orbifold CFT fixed point.    

In the ultraviolet Yang-Mills theory, there is correspondingly
a scalar potential $V(Y^I)$, where $Y^I$ are gauge-invariant
operators parameterizing the classical moduli space.
In the model inherited from the orbifold in string theory,
this potential is zero classically.  (In the quantum
field theory one could consider any 
$V(Y^I)$ classically, and we will use this freedom shortly.)  
There is an infinite number of possible terms
in $V(Y^I)$ that are not charged under any symmetry of
the orbifold theory, which therefore could be generated under
renormalization group flow.  Since for large $|Y|$ the
D1-branes separate and only feel the supersymmetry-breaking
through stretched strings of mass proportional to $|Y|$,
the potential should fall to zero as $|Y|\to\infty$.\foot{
One finds that logarithmic terms
do not arise because of cancellations in the planar
diagrams reviewed below.}
In the next section, we will explicitly calculate 
quantum corrections in the Yang-Mills regime and find
that a nontrivial potential $V(Y^I)$ is in fact generated
at subleading orders in a $1/N$ expansion.
    
From \untwisted\ we see that only a subset of these
terms become relevant operators from the point of view
of the IR orbifold conformal field theory.  
It is therefore possible to adjust parameters in
the UV Yang-Mills Lagrangian to cancel off the contributions
that will become relevant in the infrared.  In the
case of the noncompact target space \higgsmod, this
set of relevant perturbations, though a subset of the
perturbations that can be considered in the Yang-Mills
theory, is still a continuous infinity of perturbations,
and the fine-tuning we are discussing would need to be
applied to an infinite number of coefficients.  This
can be avoided by compactifying the target space.         

One could also generate corrections to the metric
on the target space \higgsmod.  In the orbifold
CFT such corrections, which can be Fourier-expanded
to the form $\partial Y\partial Y e^{ikY}$, are all irrelevant.
Therefore these terms do not need to be fine-tuned away
in order to flow to the orbifold conformal field theory.  

In the orbifold CFT, there are relevant operators in 
the $\IZ_3$ twisted sectors.  These are charged under
the $\IZ_3$ quantum symmetry of the orbifold, which,
as mentioned above, exists also in the UV Yang-Mills
theory for $\eta$ given by \ans.  
In the UV theory the operators charged under
the $\IZ_3$ symmetry cannot be generated under RG flow.

There are other symmetries of the orbifold CFT that
are not mirrored in the gauge theory.  The orbifold
CFT has a (2,2) superconformal algebra.  This algebra
includes left and right-moving $U(1)_R$ symmetries
under which some operators (twist fields) have fractional
charges, so that there is no spacetime supersymmetry.  
There is no such chiral symmetry in the Yang-Mills
quiver theories, as we discussed in \S3.1.   

Despite the appearance of an ``accidental'' supersymmetry
algebra in the IR which has no counterpart in the UV
Yang-Mills theory,  there are many features that
do agree between the two theories.
The identification of the targets space, discrete
symmetries, and untwisted
relevant perturbations
suggests strongly that the quiver Yang-Mills theory
is fine-tunable to flow to the corresponding orbifold CFT in the
infrared.

\subsec{Applications to String Theory and Gravity}

Given a well-defined (if fine-tuned) flow from the
Yang-Mills theory to the orbifold CFT, it is intriguing
to consider whether it defines an interesting gravity dual. 
At least naively, one can fine-tune the boundary conditions in
(an orbifold of) the gravity solution of \IMSY\ to
obtain an ultraviolet Lagrangian with the right
values of the couplings to flow to the orbifold CFT.  
It will be interesting to try to work this out.
Similarly we can consider the question of whether this
flow defines a matrix string theory which formulates
a spacetime string theory away from its $g_s\to 0$ limit.

Another interesting application is the following.
If we consider the $N=1$ theory, the orbifold CFT lives
simply on $\IC/\IZ_3\times \IR^6$, and constitutes a CFT
that could appear on the worldsheet of a perturbative
string.  In this context, the relevant operator in the
twisted sector is a vertex operator for a twisted-sector
tachyon in spacetime.  This is a twist field, an operator
nonlocal with respect to the elementary degrees of freedom
of the orbifold sigma model.  In the UV Yang-Mills theory,
the corresponding operator (whatever it is) is a simple
gauge-invariant combination of elementary fields in
the Yang-Mills theory.  Examples of ``twisted'' operators
in the Yang-Mills theory include relative gauge couplings
\eqn\relativegauge{
\hbox{Tr}\, F_1^2- \hbox{Tr}\, F_2^2  
}
and for example couplings of the form
\eqn\Drel{
a\left|X^1_{(+-0)}\right|^2 \, + \, b\left|X^1_{(0+-)}\right|^2
}
(analogues of relative D terms in supersymmetric theories).
Here the subscripts denote the components of $X^1$ 
involved via their charges under the $U(1)^3$ gauge symmetry.
It would be very nice
to identify the operator that corresponds to the 
vertex operators for the twisted
sector tachyons, since the effect of adding them to
the Lagrangian might be clearer in the Yang-Mills theory.    
In particular, we might get information on how much 
central charge is lost upon turning on the relevant
deformation corresponding to tachyon condensation
\oldjoe; this type of tachyon (localized at fixed
points of an orbifold) was not covered by the
techniques in \kkstach, where other types of closed
string tachyons were analyzed.  

The absence of the $U(1)_R$ chiral symmetry
in the UV theory has implications for matrix string
theory (with any amount of supersymmetry).  
One might have hoped to find that the worldsheet
supersymmetry of the infrared $g_s\to 0$ conformal
field theory would extend to the full $g_s\ne 0$ matrix
string theory.  In the infrared this symmetry is made
manifest by bosonizing the Green-Schwarz variables and
re-fermionizing them to obtain the RNS description
of the superstring where worldsheet supersymmetry is
apparent.  Given a chiral $U(1)$ current in the UV
with the right properties to match onto the $U(1)_R$
symmetry of the IR superconformal algebra, one might have
been able to bosonize that current away from criticality
and find some sort of RNS formulation of the matrix
string theory.  This does not seem possible for
the quiver theories (or the parent theory of
D1-branes in flat space); it would be interesting to
find backgrounds for which such chiral symmetries
{\it would} appear on the worldvolume of D1-branes.

\newsec{Diagrammatics}
       
\medskip

In this section we perform explicit calculations to check
whether the terms of the form $V(|Y|)$ are in fact generated
under RG flow in the Yang-Mills regime ($g_{YM}$ small)
of the quiver theory.  
After reviewing the fact that they are not generated
by planar diagrams, we show that they are generated by
nonplanar diagrams.  By way of contrast, we show that in the 
asymmetric orbifold case of \S2.3, such terms are
{\it not} generated at the order we compute.  
For simplicity in this section we will work with the
$N=1$ theory of a single D1-brane on the orbifold background.

\subsec{Planar Diagrams}

The quiver quantum field theory has a dual description
in terms of an orbifold of the gravity solution of \IMSY,
which acts on the sphere surrounding the branes but not
on the radial coordinate \ksorb.  In the worldsheet
string theory describing this gravity background,
correlation functions of untwisted vertex operators
are inherited from those of the unorbifolded theory.  
This ensures that
the correlation functions of untwisted operators in
the gauge theory are similarly inherited, if the
correspondence is correct (as explained
for the $\beta$-functions in \ksorb).  This result can
be seen directly to all orders in perturbation theory
by an analysis of the field theory diagrams \bkv.
  
So for example in the case of the 1-loop contributions
to the scalar mass squared for $X^1$ in the $\IC/\IZ_3$ quiver
theory, one finds the following graphs:
%%%{\it !!!Planar 1-loop graphs figure}

%%%%%%%%%%%%%%%%%%%%%%% Figure 3 %%%%%%%%%%%%%%%%%%%%%%%%%%%%%%%%%
\smallskip
\bigskip
\centerline{\vbox{\hsize=4.5in\tenpoint 
\centerline{\epsfbox{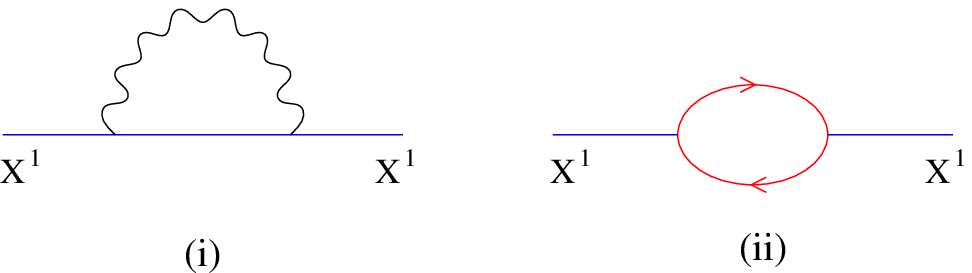}}
\smallskip\noindent
Figure 3: Planar 1-loop Feynman diagrams contributing to the $X^1$ mass
term in the $\IC/\IZ_3$ quiver theory.}}
\bigskip
%%%%%%%%%%%%%%%%%%%%%%%%%%%%%%%%%%%%%%%%%%%%%%%%%%%%%%%%%%%%%%%%%%

The diagram (i) with bosons running in the loop cancels
against a diagram (ii) with fermions running in the loop;
although the fermions are not superpartners of the bosons
they come in the right multiplicities to cancel and the Lorentz
structure of the diagrams are the same as appears in the
supersymmetric theory.

\subsec{Non-planar diagrams}

Let us first consider non-planar contributions to
the scalar potential on the Coulomb branch.
Let us calculate the mass squared of the modes describing
the relative motion of the branes in the $X^{2,3,4}$
directions.  At one-loop the relevant diagrams are
%%{\it !!! Figure:  1-loop Coulomb branch}

%%%%%%%%%%%%%%%%%%%%%%% Figure 4 %%%%%%%%%%%%%%%%%%%%%%%%%%%%%%%%%
\bigskip
\centerline{\vbox{\hsize=4.5in\tenpoint
\centerline{\epsfbox{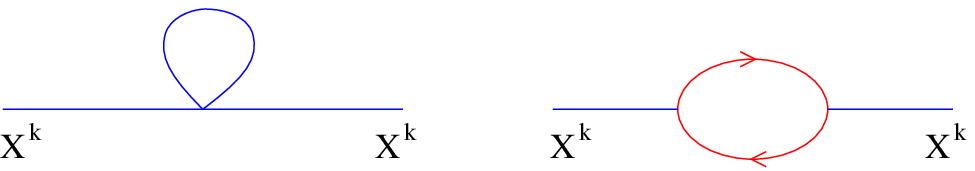}}
\smallskip\noindent
Figure 4: 1-loop contribution to the scalar potential on
the Coulomb branch. The index $k=2,3$ or 4.}}
\bigskip
%%%%%%%%%%%%%%%%%%%%%%%%%%%%%%%%%%%%%%%%%%%%%%%%%%%%%%%%%%%%%%%%%%

This includes nonplanar contributions.  The zero-external-momentum
piece of this diagram does not cancel.  This can
be seen from the fact that there is an excess of
fermions running in the loop as compared to the balance
of fermions and bosons in corresponding diagrams in a
supersymmetric theory.  

On the Higgs branch the relevant quantity to compute is
the scalar potential for $X^1$.  Here there are no
non-planar contributions at one loop.  At two loops
we find nonplanar diagrams of the form 
%%{\it !!!figure: two-loop diagrams}.  

%%%%%%%%%%%%%%%%%%%%%%% Figure 5 %%%%%%%%%%%%%%%%%%%%%%%%%%%%%%%%%
\bigskip
\centerline{\vbox{\hsize=4.5in\tenpoint
\centerline{\epsfbox{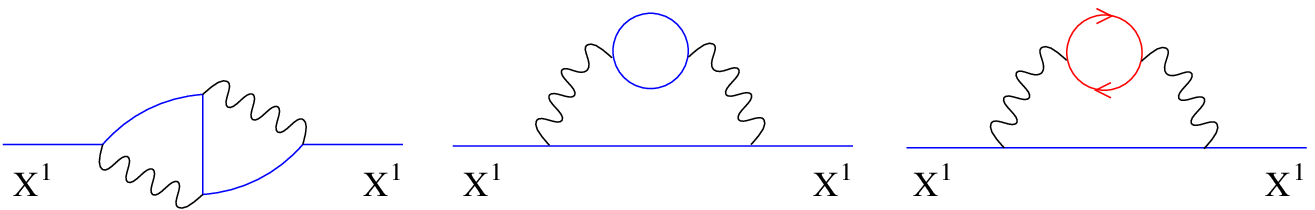}}
\smallskip\noindent
Figure 5: 2-loop contributions to the scalar potential on
the Higgs branch.}}
\bigskip
%%%%%%%%%%%%%%%%%%%%%%%%%%%%%%%%%%%%%%%%%%%%%%%%%%%%%%%%%%%%%%%%%%

In a supersymmetric theory, their contribution
to the scalar potential would cancel against the contribution
of the following graphs involving the superpartners.
%%%{\it !!!figure:  2-loop superpartner diagrams}  

%%%%%%%%%%%%%%%%%%%%%%% Figure 6 %%%%%%%%%%%%%%%%%%%%%%%%%%%%%%%%%
\bigskip
\centerline{\vbox{\hsize=4.5in\tenpoint
\centerline{\epsfbox{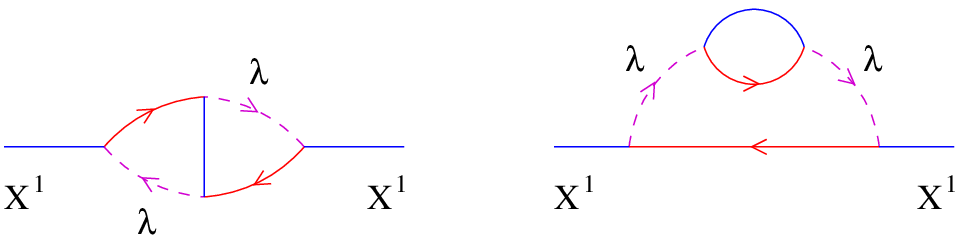}}
\smallskip\noindent
Figure 6: 2-loop ``superpartner'' Feynman diagrams whose
contributions  would cancel against those in Figure 5.  Here $\lambda$
denotes gauginos.}}  
\bigskip
%%%%%%%%%%%%%%%%%%%%%%%%%%%%%%%%%%%%%%%%%%%%%%%%%%%%%%%%%%%%%%%%%%
\noindent
These fermions running in the loops of
these diagrams include adjoint fermions (gauginos in a 
supersymmetric theory).  Evaluating these ``superpartner''
diagrams, we find for the first one
\eqn\diagone{\eqalign{
&(-ig_{YM})^4\int d^2q d^2k {{tr ((\sslash{k}+m)\sslash{q}
(\sslash{q}+m) \sslash{k})}
\over{(q^2+m^2)q^2(k^2+m^2)k^2(q-k)^2}}\cr
&=(-ig_{YM})^4\int d^2q d^2k {{k^2q^2+m^2q\cdot k}
\over{(q^2+m^2)q^2(k^2+m^2)k^2(q-k)^2}}\cr
}
}  
For the second one we find
\eqn\diagtwo{\eqalign{
&(-ig_{YM})^4\int d^2q d^2k {{tr 
((\sslash{k}+m)\sslash q(\sslash{k}+m) \sslash{k})}
\over{q^2(k^2+m^2)^2k^2(q-k)^2}}\cr
&=(-ig_{YM})^4\int d^2q d^2k {{(k\cdot q)(k^2+m^2) }\over
{(k^2+m^2)^2k^2q^2(q-k)^2}}\cr
}
}
Here the mass $m$ for the gauginos is proportional
to $g_{YM}\rho$, where $\rho$ is the distance of
the D1-brane from the orbifold fixed point.  
Let us consider the integrands of these diagrams.
We find, upon introducing Feynman parameters, an
expression that separates into a piece that integrates
to zero plus a piece whose integrand is of definite
sign that cannot cancel.  To begin with the
$k\cdot q$ terms in \diagone\diagtwo\ are
of indefinite sign.      
Introducing Feynman parameters, one obtains an integrand
of the form 
\eqn\genform{
\left( {{f(k^2,m^2)}\over D((q^\prime)^2, k^2)} \right)^5 k\cdot(q^\prime + ak) 
}
for each of these contributions.  Here the denominator $D$ 
is a positive (indefinite) function of $(q^\prime)^2$
and $k^2$, where $q^\prime$ is equal to $q-ak$ for
a positive quantity $a$ (which is a function of Feynman
parameters); $f(k^2,m^2)$ is a positive
function of $k^2$ and $m^2$.  Given this, the 
$k\cdot q^\prime$ term in the numerator integrates
to zero since the integrand is odd in $q^\prime$
and $k$.  The other term $ak^2$ has a positive integrand.
These surviving contributions to the integrand from
the $k\cdot q$ terms have the same sign as the other
terms in \diagone\ and \diagtwo.  

Since the surviving integrand here is of definite sign, 
the contribution of the ``superpartner'' graphs
is nonzero, independent of the precise infrared regularization
scheme that we might introduce, as long as that scheme
respects Lorentz invariance.  (So this result would hold
for example if we introduce masses to regularize the
infrared divergences.)    

This above non-cancellation of 1/N corrections to the
D1 worldvolume theory will occur for generic nonsupersymmetric
quiver theories.  The simplest way to see this
is to note that the absence of
adjoint fermions (which would run in the loops of
\diagone\ and \diagtwo) is a direct consequence of the fact that
the orbifold action projects out all spacetime spinors.
Given a nontrivial action on the spinor quantum numbers
of the worldvolume fermions, the surviving representations
are necessarily bifundamental rather than adjoint.  

\subsec{Asymmetric Orbifolds}

This is not the case for the asymmetric orbifolds reviewed
briefly in \S2.3.  For these theories, the invariant
states of the form $\left|\phi\right>_{1-1}+\left|g\phi\right>_{5-5}$ include
adjoint fermions.  Their interactions are also such
that the ``superpartner'' diagrams \diagone\diagtwo\
are present in the theory, in addition to the bosonic
diagrams that they cancel.  As such, at this order
in the asymmetric orbifold background, the moduli
space on the D-brane worldvolume is stable under
quantum corrections.  These diagrams are suppressed
by a factor of $1/N^2$ relative to the leading order contributions.
This cancellation is in accord with the
1-loop cancellation of the cosmological constant
in this background \kks\ through AdS/CFT duality
as explained in \ksorb.  It would be very interesting
to see where the cancellations break down (if
at all) from the point of view of the D-brane worldvolume
theories.  In particular, the spacetime theory's 2-loop
cancellation in the cosmological constant 
(according to a rather subtle calculation in \kks) suggests
that these cancellations might persist at least to order
$1/N^4$ in the D-brane theories.\foot{This two-loop
cancellation is based on the rather subtle formalism
of integration over split RNS supermoduli space at
genus two, evaluated in a particular gauge with separate
analysis of the cancellation of boundary contributions.
A recent work \iz\ objecting to this
conclusion for a subset of these theories 
includes a not-yet-complete calculation
in a different gauge, and (as far as we understand)
an invalid objection to
a determinant factor in the measure of the integral
over supermoduli space included in the calculation
of \kks.  Independent tests of the result
of the calculation in \kks\ (and therefore of
the formalism involved) will involve integration over the
moduli space of the diagram; such tests might
be most tractable near boundaries of the moduli space \vafa.}       

\newsec{Singular CFTs}

\medskip

As discussed above, for $\theta=0$ a new (Coulomb) branch opens
up in the target space of the Yang-Mills theory.  
In supersymmetric situations, the existence of this branch is stable
against quantum corrections (although its geometry is 
subject to corrections).  As first explained in the context
of Calabi-Yau compactification in \phases, these branches
lead to singularities in correlation functions arising
from integration over the bosonic zero modes parameterizing
this branch \silvwitt.  In the application to worldsheet
string theory, these singularities in the 
infrared CFT describe      
spacetime backgrounds of string theory in which light
non-perturbative states appear (and resolve the singularities
in the correlation functions).  In particular
in compactifications of type II string theory to six
dimensions on $K3$, one finds non-perturbatively
enhanced non-abelian gauge symmetry at points 
in moduli space where the geometry is an orbifold
space (an ADE singularity) and the integrals 
$\int_{C_i} B$ of the NS 2-form
$B$ over the collapsed 2-cycles $C_i$ of the ADE singularity
vanish \ade.  
The quantum-mechanical explanation of the singularity
is different in different supersymmetry classes 
\conifold.
It is interesting
to consider whether this kind of phenomenon arises also
in non-supersymmetric string backgrounds.  

In generic non-supersymmetric orbifold conformal field theory
(for example on the $\IC/\IZ_3$ background we have been
considering) one does not expect exactly marginal operators
like $\int B$ of the above example.  Let us instead consider
an orbifold conformal field theory on the product space
$\IC/\IZ_3\times \IC^2/\IZ_2$.  The second factor alone preserves
supersymmetry and does have exactly marginal operator in
the $\IZ_2$ twisted sector corresponding to $\int B$.  
This exactly marginal operator is trivially present in
the product as well, and when $\int B\to 0$ the correlation
functions of the full theory (in particular those involving
$\IZ_3$-invariant, $\IZ_3$-untwisted operators) diverge
as in the supersymmetric theory.  

This singularity in the non-supersymmetric worldsheet
conformal field theory as $\int B\to 0$ 
will not cancel upon considering condensation of 
the $\IZ_3$-sector tachyon and higher-loop effects.  
It suggests the presence of light states in the theory not
described by fundamental strings.  It would be interesting
to understand the resolution of this singularity in
the spacetime theory.  Given such non-supersymmetric backgrounds
not described by perturbative string theory, it
would be interesting to understand the contributions
these new light states make to quantum effects in
these backgrounds.    

We can construct the quiver theory corresponding to this
product orbifold CFT.  One finds a $U(1)^6$ gauge group
and matter content given by the following quiver diagrams:
%%%{\it !!!quiver diagrams for product theory}   

%%%%%%%%%%%%%%%%%%%%%%% Figure 7 %%%%%%%%%%%%%%%%%%%%%%%%%%%%%%%%%
\bigskip
\bigskip
\centerline{\vbox{\hsize=4.5in\tenpoint
\centerline{\epsfbox{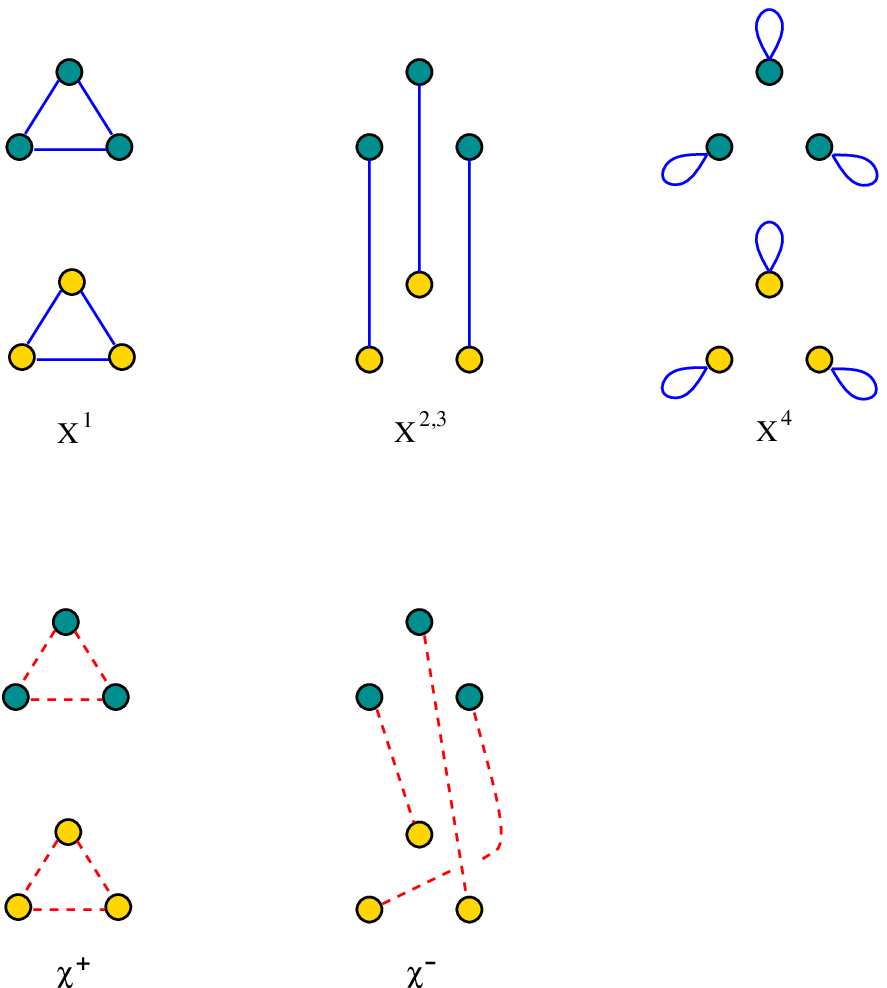}}
\smallskip\noindent
Figure 7: $\IC/\IZ_3\times \IC^2/\IZ_2$ quiver Diagrams. $\chi^+$
($\chi^{-}$) collectively denote the spinors with $+1 (-1)$
eigenvalues under the $\IZ_2$ action on $\IC^2$.} }
\bigskip
%%%%%%%%%%%%%%%%%%%%%%%%%%%%%%%%%%%%%%%%%%%%%%%%%%%%%%%%%%%%%%%%%%

Note that in the Yang-Mills theory, all degrees of freedom
mix with each other and there is no product structure.
In the quiver theory there is a branch of classical
moduli space in which fractional branes separate at
the $\IZ_2$ fixed point, when the relative $\theta$
angle $\theta_{\IZ_2}$ of the $\IZ_2$ part of the 
quiver theory vanishes.  

When the $\theta$ angles of the UV theory satisfy
the discrete symmetry constraints of \S3.1,
the theory can be fine-tuned to flow to the orbifold
CFT on $\IC/\IZ_3\times \IC^2/\IZ_2$.  It would be interesting
to understand whether this feature persists as we take
$\theta_{\IZ_2}$ to zero (in particular one would
like to know how much fine-tuning
is required as a function of $\theta_{\IZ_2}$).    
If it persists, then one could study the singularity via
the Coulomb branch of the Yang-Mills theory as in other cases.

In \donedfivecft, the singular CFT in (4,4) supersymmetric theories
was studied, and an effective description in terms of
a WZW model times a linear dilaton was obtained far
down the throat of the target space.  As in the case
of orbifold conformal field theory, this type of CFT
exists (and is exactly solvable) with arbitrary amounts
of supersymmetry.  It would be interesting to find
examples of theories with broken supersymmetry where
this type of ``throat'' theory emerges.  

Unlike in the case of orbifold CFT, the WZW times linear
dilaton part of the singular CFT describes only a piece of
the target space, which must be matched on to the rest
of the target space in a manner which is consistent with
conformal invariance.  One might hope to access
conformally invariant theories of this type by
arriving at them via marginal deformation from an 
understood conformal field theory, such as orbifold
conformal field theory.  Indeed, in the supersymmetric
context one finds a flow pattern of the following form:
%%{\it !!!Figure describing YM->IR flow for theta 0 and pi}

%%%%%%%%%%%%%%%%%%%%%%% Figure 8 %%%%%%%%%%%%%%%%%%%%%%%%%%%%%%%%%
\bigskip
\centerline{\vbox{\hsize=4.5in\tenpoint
\centerline{\epsfbox{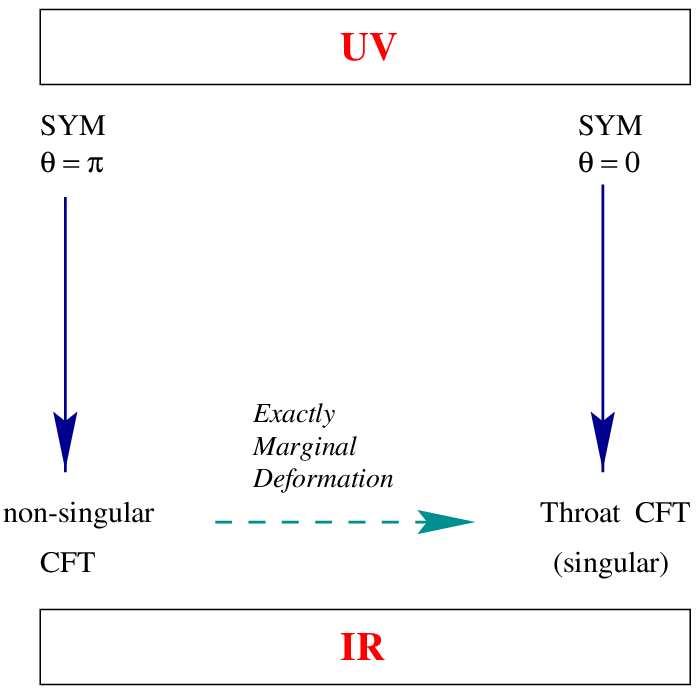}}
\smallskip\noindent
Figure 8: An expected flow pattern in the presence of supersymmetry.}}
\bigskip
%%%%%%%%%%%%%%%%%%%%%%%%%%%%%%%%%%%%%%%%%%%%%%%%%%%%%%%%%%%%%%%%%%

For example, in the D1-D5 system with $Q_1=1$ and $Q_5=2$,
the Higgs branch moduli space is $\IR^4/\IZ_2\times \IR^4$.
There are four exactly marginal perturbations involving the twist
fields for the $\IZ_2$ action.  
In the D1-D5-system with a transverse $\IC/\IZ_3$ orbifold
action described in \S2.2\ (for $Q_1=1,Q_5=2$), we find a 
Higgs branch moduli 
space 
\eqn\modd{
{\cal M}=(\IR^4/\IZ_2)^3\times \IR^4
}
Unfortunately in this case, we find that the $\IR^4/\IZ_2$ twist
fields have dimension greater than 1 (and are therefore
irrelevant).  This arises because there are more fermions
transforming under the relevant $\IZ_2$ factor in the
Yang-Mills gauge group than in the corresponding supersymmetric
theory.  So here we cannot deform toward a singular CFT
using a marginal operator, since none presents itself.
It would be very interesting to understand whether
nonetheless the $\theta=0$ Yang-Mills theory flows
to a nontrivial (singular) CFT in the IR directly,
perhaps with fine-tuning.

\centerline{\bf Acknowledgements}

We would like to thank O. Aharony, T. Banks, M. Berkooz, M. Douglas,
K. Intriligator, S. Kachru, A. Kapustin, A. Karch,   
N. Seiberg, S. Shenker, M Strassler, 
C. Vafa, and E. Witten for very helpful
discussions about various aspects of this project.  We would
like to thank the Institute for Advanced Study for hospitality
during the bulk of this work.  The work of E.S. is supported by
the DOE by an OJI grant and under contract
DE-AC03-76SF00515, and by the A.P. Sloan Foundation.  The
work of Y.S. is supported by an NSF graduate fellowship.      
%%%%%%%  NSF Graduate Fellowship, DOE, etc.

\listrefs

\end